\documentclass[12pt]{iopart}
\usepackage{graphicx,epsfig}
\begin{document}

\title[]{Memory effects in vibrated granular systems}

\author{J. Javier Brey \footnote[1]{ brey@cica.es} and A. Prados
\footnote[2]{prados@cica.es}}

\address{F\'{\i}sica Te\'{o}rica. Universidad de Sevilla. Apartado de Correos
1065. 41080 Seviila. Spain}

\begin{abstract}
Granular materials present memory effects when submitted to tapping
processes. These effects have been observed experimentally and are discussed
here in the context of a general kind of model systems for compaction
formulated at a mesoscopic level. The theoretical predictions qualitatively
agree with the experimental results. As an example, a particular simple model is
used for detailed calculations.

\end{abstract}



\maketitle

\section{Introduction}
\label{s1}
A granular material is a system composed of a ``large'' number of macroscopic
particles or grains whose interactions are inelastic, so that mechanical energy
is not conserved. Here, in practice, large means in many cases a few
thousands or even hundreds of particles, as opposite to usual molecular systems
having a number of particles of the order of Avogadro's number. The
macroscopic size of the
particles implies that the usual concepts of thermodynamics can not be directly
translated to these systems. For instance, if we consider a typical granular
system such as sand, the energy needed to lift a grain by one diameter is more
than ten orders of magnitude larger than the thermal energy of the grain at
room temperature. Moreover, due to the inelasticity of collisions it is
necessary to supply external (nonthermal) energy to the system in order to
generate a steady state or to study the relaxation of the system towards
a stable configuration.
Therefore, although the concept of {\em granular temperature}
is often used in the literature, it must be understood just as a measure of
the velocity fluctuations in the system, without being related with any
underlying idea of thermal equilibrium.

The phenomenology of granular media is very rich, showing many characteristic
complex behaviours \cite{JNyB96}. Here we will focus in one them,  namely
compaction, that can be roughly defined as the nonthermal relaxation of a
loosely packed system of many grains under vertical mechanical tapping or
vibration. This problem is of fundamental importance to many industrial
applications and also raises fundamental theoretical questions. Therefore, it is
not surprising that in the last years systematic experimental investigations and
theoretical approaches are being developed to described the dynamics of
compaction processes, as well as the nature of the final state reached by the
system.

\begin{figure}
\centerline{\includegraphics[scale=0.5]{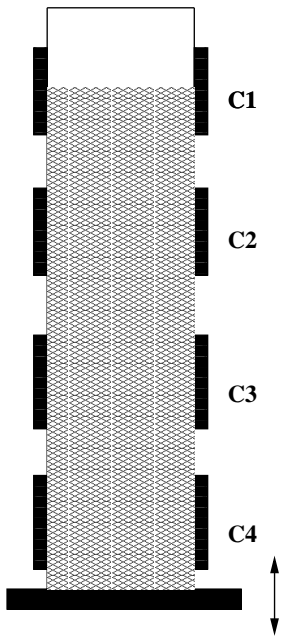}}
\caption{\label{f1}Sketch of the experimental set-up used in Reference
\protect{\cite{KFLJyN95}}. The density is measured at four different heights
by means of the capacitors $C1-C4$.}
\end{figure}

The evolution of the density in vibrated granular materials was investigated
in a pioneering paper by Knight
\etal \cite{KFLJyN95,NKBJyN98}. The experimental set-up they used consisted
of monodisperse spherical glass beads in a long thin cylinder mounted vertically
on a vibration exciter. The packing fraction of the beads was measured by means
of four capacitors mounted at different heights along the tube (see Figure
\ref{f1}). The shaking intensity $\Gamma$, determined by the  maximum applied
acceleration normalized by gravity, was controlled. The system was prepared in a
low density initial state before being submitted to a sequence of single shakes
or ``taps''. Time between taps was long enough to allow the system to come to
rest, so they were completely independent and internal resonances were avoided.
It was observed that the density increased monotonically, tending eventually to
a steady value. The number of taps required to reach the final density was very
large, often larger than $10^{5}$.

The authors found that a good description of the experimental data was obtained
with an inverse logarithm four parameter fit of the form
\begin{equation}
\label{1.1}
\rho(t)=\rho_{\infty}-\frac{\Delta \rho_{\infty}}{1+B \ln (1+t/t_{c})}\, ,
\end{equation}
where time is measured in number of taps and  the parameters $\rho_{\infty}$,
$\Delta \rho_{\infty}$, $B$, and $t_{c}$ are constants that depend only on
the tapping strength $\Gamma$. The above inverse logarithmic expression fitted the
data better
than other more usual laws such as a single exponential, a combination of
two exponentials, a power law, or a stretched exponential. Two features
of the experimental results that are relevant for the later discussion here
are
\begin{enumerate}
\item The slope of the relaxation curve is smaller for smaller vibration
intensity $\Gamma$, i.e. the relaxation is slower for smaller $\Gamma$.
\item The final steady density is a monotonic decreasing
function of $\Gamma$ \cite{note1}.
\end{enumerate}
A theoretical attempt to formulate a ``thermodynamic'' description of the
steady states reached by the system in the long time limit has been carried
out by Edwards and coworkers \cite{EyO89,MyE89,EyM94}. They extended the
methods and concepts of usual statistical physics to granular media.
The basic idea is that the volume plays in these systems a role analogous
to that played by the energy in molecular (elastic) systems. Although
there is no experimental verification of this theory up to now, it has been
shown to be consistent with the behaviour of some simple models for
granular compaction \cite{BPyS00a,LyD01,BKLyS00}.

In addition to the slow relaxation described above, vibrated granular
materials exhibit a series of properties that are reminiscent of the
typical behaviour of conventional structural glasses. This includes effects
as annealing, i.e. slow ``cooling'' properties, and hysteresis, when submitted
to processes in which the tapping intensity is monotonically increased
and decreased \cite{NKBJyN98,Ja97}. This apparent glassy nature of granular
compaction led Josserand \etal \cite{JTMyJ00} to investigate the response of a
vibrated granular system to sudden perturbations of the intensity $\Gamma$.
Their work was suggested by classical experiments for the study of aging in
glasses, and the realization that the vibration intensity plays in vibrated
granular media a role analogous to the temperature of molecular systems.
Previously, a similar process had been studied by means of numerical
simulations in a model for compaction \cite{Ni99}.

In the simplest compaction experiment of Reference \cite{JTMyJ00}, the vibration
intensity was instantaneously changed from a value $\Gamma_{1}$ to another
$\Gamma_{2}$ after $t_{w}$ taps. For $\Gamma_{1} > \Gamma_{2}$ it was observed
that on short times scales the compaction rate increases, while for $\Gamma_{2}
> \Gamma_{1}$ the system dilates for short times. Both results are opposite
from the behaviour at constant $\Gamma$, as discussed above. After several
taps, the
``normal'' behaviour was recovered. This is a direct evidence of the
presence of short-term memory effects in the system, so that the density for
$t>t_{w}$ is not determined by its value at $t=t_{w}$. Other experiments
are also reported by the authors showing the same kind of non-Markovian
behaviour. In the next section we will investigate whether this anomalous
response can be understood in terms of simple and general arguments.

\section{Mesoscopic description of the density evolution}
\label{s2}
When a granular system is being vibrated, different kind of
events  take place in the system. The existence of a steady density for
constant vibration intensity suggests the presence of two different kinds
of elementary processes, the ones trying to increase the density and the
others trying to decrease it. Stationarity arises when both tendencies
cancel each other. Then, we will assume that the time evolution
of the density $\rho$ in a tapping process can be described at a mesoscopic
level by a balance equation of the form
\begin{equation}
\label{2.1}
\frac{d \rho (t)}{dt}=f_{1}(\Gamma) \mu_{1}(t)-f_{2}(\Gamma)\mu_{2}(t),
\end{equation}
where $t$ is measured in number of taps, and the several functions verify
\begin{equation}
\label{2.2}
f_{1}(\Gamma) \geq 0, \quad f_{2}(\Gamma) \geq 0,
\end{equation}
\begin{equation}
\label{2.3}
\mu_{1}(t)>0, \quad \mu_{2}(t) >0.
\end{equation}
The quantities $\mu_{1}(t)$ and $\mu_{2}(t)$ are assumed  to include all the
correlation effects. Note that we do not assume that $\mu_{1}$ and $\mu_{2}$ can
be expressed as functions of $\rho(t)$, so that Equation (\ref{2.1}) is not
closed. Our aim in the following will be to investigate what information can be
obtained about $f_{1}$, $f_{2}$, $\mu_{1}$, and $\mu_{2}$ by using very general
and plausible physical arguments. First, we notice that since we are measuring
the time in units of complete taps, if there is no tapping there is no time
evolution either. Therefore, it must be
\begin{equation}
\label{2.4}
f_{1}(\Gamma=0)=f_{2}(\Gamma=0)=0.
\end{equation}
Moreover, the number of elementary processes, both tending to increase and to
decrease the density, are expected to increase with $\Gamma$. This
expectation is reinforced by the analogy of $\Gamma$ with the temperature,
as mentioned in the previous section. Then we assume that $f_{1}(\Gamma)$
and $f_{2}(\Gamma)$ are both monotonic increasing functions.
Let us introduce the function
\begin{equation}
\label{2.6}
g(\Gamma)\equiv \frac{f_{2}(\Gamma)}{f_{1}(\Gamma)}\, ,
\end{equation}
representing the ratio of the rate associated to the decompaction processes to
that of the compaction ones. As already mentioned, experiments show that the
steady density is a decreasing function of $\Gamma$. This suggests that
$g(\Gamma)$ must be an increasing function of the vibration intensity, so that
decompaction events become relatively more relevant. This assumption can be
further justified by analysing the behaviour of the steady solutions of Equation
(\ref{2.1}) \cite{ByP01a}. Then we can rewrite our mesoscopic evolution equation
as
\begin{equation}
\label{2.7}
\frac{d}{dt} \rho(t)=f_{1}(\Gamma) \left[ \mu_{1}(t)-g(\Gamma) \mu_{2}(t)
\right].
\end{equation}
Here $f_{1}(\Gamma)$, $g(\Gamma)$, $\mu_{1}(t)$, and $\mu_{2}(t)$ are all
positive quantities. Moreover, the first two ones are increasing functions
of $\Gamma$, vanishing in the limit $\Gamma \rightarrow 0$. This fully
specifies the general kind of evolution equations we will deal with.
Let us consider the following experiment. A system is vibrated
with a constant intensity $\Gamma$. At a given time $t_{w}$, the intensity
is instantaneously changed to $\Gamma + \Delta \Gamma$. We want to study the
change in the relaxation rate $r(t) \equiv d \rho (t) /dt$. Just {\em before}
the change we get from (\ref{2.7}),
\begin{equation}
\label{2.8}
r_{w}=f_{1}(\Gamma) \left[ \mu_{1}(t_{w}^{-})-g(\Gamma) \mu_{2} (t_{w}^{-})
\right],
\end{equation}
while just {\em after} the change it is
\begin{equation}
\label{2.9}
r_{w}^{\prime}=f_{1}(\Gamma+\Delta \Gamma) \left[ \mu_{1}(t_{w}^{+})-
g(\Gamma+\Delta \Gamma) \mu_{2} (t_{w}^{+}) \right].
\end{equation}
Then, for $\Delta \Gamma \ll 1$, we get
\begin{equation}
\label{2.10}
\frac{\Delta r_{w}}{\Delta \Gamma} \equiv \frac{r^{\prime}_{w}-r_{w}}{\Delta
\Gamma} = \lambda (t_{w}),
\end{equation}
where we have defined the function ($r(t)$ and $\mu_2(t)$ are computed over the
relaxation curve at constant $\Gamma$)
\begin{equation}
\label{2.11}
\lambda(t)=\frac{f^{\prime}_{1}(\Gamma)}{f_{1}(\Gamma)} r(t)-f_{1}(\Gamma)
g^{\prime} (\Gamma) \mu_{2}(t).
\end{equation}
Upon deriving Eq.\ (\ref{2.10}) we have assumed that $\mu_1(t)$ and $\mu_2(t)$
are continuous at $t=t_w$. This seems physically plausible since they are
functionals of the state of the system. For instance, in the mesoscopic
description they can be expressed as given  moments of the underlying
distribution function of the system, which is known to be a continuous function
of time. In the limit $t_{w}\rightarrow\infty$, the system has time to reach the
stationary density corresponding to the intensity $\Gamma$ before the change, so
that $r_{w}=0$ and $\lambda (t_{w})<0$, because of the properties of the
functions $f_{1}(\Gamma)$ and $g(\Gamma)$. In the opposite limit, $t_{w}
\rightarrow 0$, and assuming that the initial density is close to its minimum
value, it is $\lambda (t_{w}) >0$. To derive this, we have taken into account
that $\mu_{2}$ must vanish in the lowest density limit, where by definition no
processes decreasing the density are possible. The conclusion of this discussion
is that $\lambda (t_{w})$ has opposite signs in the short and large $t_{w}$
limits. Then, it follows from Eq.\ (\ref{2.10}) that when the system has been
vibrated for a short time before the change of vibration intensity, $\Delta
r_{w}$ and $\Delta \Gamma$ have the same sign, i.e. the response of the system
is what we have described in the previous section as normal. On the other hand,
if the vibration time before the change is large, the response of the system is
anomalous. In this context, the experiments reported in Reference \cite{JTMyJ00}
would correspond to ``large'' vibrating periods before the abrupt change in the
shaking intensity.

\section{A simple model for granular compaction}
\label{s3}
In this Section we are going to describe a one-dimensional model for granular
compaction \cite{BPyS99} that is simple enough as to allow for detailed
calculations,
and at the same time captures many of the experimental characteristic
features. We consider a lattice in which each site $i$ can be either occupied
by a particle or empty (occupied by a hole). A variable $m_{i}$ is assigned to
each site, taking the value $m_{i}=1$ if the site is empty, and $m_{i}=0$ if
there is a particle on it. The time evolution of the system is defined in the
following way. The only possible elementary events occurring in the
system are the adsorption of a particle on an empty site and the desorption
of a particle from the lattice to the bulk. Then the dynamics is formulated
by means of a master equation with a transition rate for the change of
$m_{i}$ into $1-m_{i}$ given by
\begin{equation}
\label{3.1}
W_{i}({\bf m})=\frac{\nu}{2} ( m_{i-1}+m_{i+1}) \left[ \epsilon+m_{i} (1-2
\epsilon) \right].
\end{equation}
Here ${\bf m} \equiv \{m_{1},m_{2},\ldots \},$ $\nu$ is a frequency defining
the characteristic time scale, and $\epsilon$
is a dimensionless parameter taking values in the interval $0 \leq \epsilon
\leq 1$. Thus the transition rate for the adsorption of a particle at site
$i$ is
\begin{equation}
\label{3.2}
W_{i}^{+}({\bf m})=\frac{\nu}{2} (1-\epsilon) m_{i} (m_{i-1}+m_{i+1}),
\end{equation}
and that for the desorption of a particle
\begin{equation}
\label{3.3}
W_{i}^{-}({\bf m})=\frac{\nu}{2} \epsilon (1- m_{i}) (m_{i-1}+m_{i+1}).
\end{equation}
It is seen that for $\epsilon=1$ no particle can be absorbed, while for
$\epsilon=0$ desorption processes are inhibited. Moreover, all processes are
restricted, in the sense that a site can change its state only if at least one
of its nearest neighbour sites is empty. A similar kind of facilitated  dynamics
has been considered in the formulation of  Ising models for glassy relaxation
\cite{FyA84}. A physical picture of the model can be obtained by associating a
hole with a region of the granular system having a low density, and a site
occupied by a particle with a high density region. The facilitated dynamics
tries to translate the idea that fluctuations leading to rearrangements in a
region require a neighbour region with low packing fraction. Our one-dimensional
model is a very idealized representation of a low horizontal layer in a vibrated
granular system, the parameter $\epsilon$ being related with the intensity of
vibration.

In order to model a tapping experiment, the system is initially placed in
a purely random configuration, from which it relaxes with $\epsilon=0$
until it gets eventually trapped in a metastable state, with all
the holes surrounded by two particles. This is a low density configuration
that is taken to correspond to the loosely initial conditions used in
real experiments. Then, a pulse is generated by instantaneously increasing
$\epsilon$ to a given value for a time period $t_{0}$.
Afterwards, the system relaxes again with no external excitation, i.e.
with $\epsilon =0$. In order to mimic what is done in experiments, the
relaxation lasts long enough as to allow the system
to reach a metastable configuration, from which no evolution is
possible in absence of external excitation. This completes a tap event. The
process is repeated to generate a series of them.

A first test of the relevance of the model to study compaction in vibrated
granular systems is, of course, whether it leads to the same kind of density
evolution at constant vibration intensity as experimentally observed. We have
verified that this is the case in the limit $\epsilon \nu t_{0} \ll 1$. The mean
density in an homogenous configuration  is given by $\rho(t) \equiv 1- \langle
m_{i}(t) \rangle$, with the angular brackets denoting ensemble average and $i$
being arbitrary.

\begin{figure}
\centerline{\includegraphics[scale=0.5]{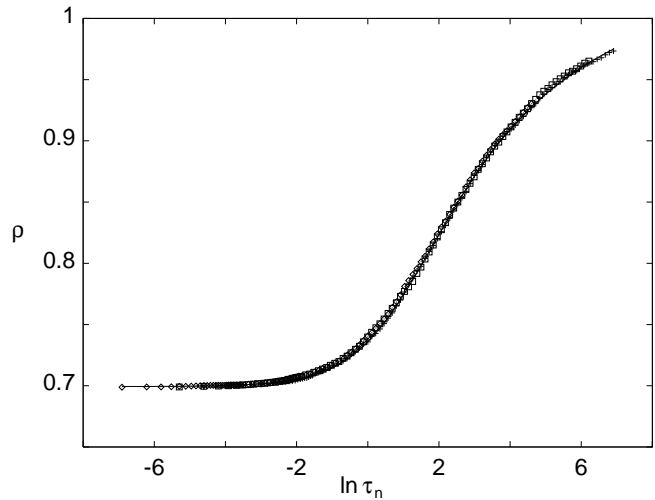}}
\caption{\label{f2}Time evolution of the density of the model described in
the main text when it is submitted to a tapping process with $\epsilon= 0.5$.
Data obtained with three different values of $t_{0}$, namely $0.002$, $0.001$,
and $0.02$ have been plotted.}
\end{figure}

An example is given in Figure \ref{f2}, where the relaxation of the density is
plotted as a function of the scaled time $\tau_{n}=\nu \epsilon t_{0} n$, $n$
being the number of taps. The numerical data have been obtained from Monte Carlo
simulations and different values of $\epsilon$ and $t_{0}$ have been used. The
fact that all curves collapse indicates that $\tau_{n}$ is the relevant time
scale for the compaction problem. Also plotted in the same figure is the fit to
the phenomenological law (\ref{1.1}), with parameters $\rho_{\infty}=1.10$,
$\Delta\rho_{\infty} =0.40$, $B=0.39$, and $t_{c}= 3.37/\nu \epsilon t_{0}$. It
is seen that the inverse logarithmic law describes very well the simulation
results. A similar behaviour has also been found in other models for granular
compaction \cite{CLHyN97,NCyH97,BKNJyN98}. Nevertheless, we ought to say that we
have not been able to derive the heuristic relaxation law by analytical methods,
in spite of the tractability of our model. It is possible that it be just a
convenient fitting expression over a wide time window. A strong indication
supporting this idea is that the steady density predicted by the logarithmic law
$\rho_{\infty}$ in the limit of an asymptotically large number of taps not only
is in disagreement with the simulations, but is clearly unphysical since it is
larger than one. The same happens with the experimental results, where
$\rho_{\infty}$ is sometimes greater than the random close packing
\cite{KFLJyN95}. In fact, it is possible to derive an analytical expression for
the asymptotic density reached by the model with the result
\begin{equation}
\label{3.4}
\rho_{0}^{(s)}\simeq1- \frac{1}{2} \epsilon \nu t_{0},
\end{equation}
valid again in the limit $\epsilon \nu t_{0} \ll 1$. Also this expression
has been checked by Monte Carlo simulations \cite{BPyS99}.

\section{Effective dynamics for tapping processes}
\label{s4}

\begin{figure}
\centerline{\includegraphics[scale=0.5]{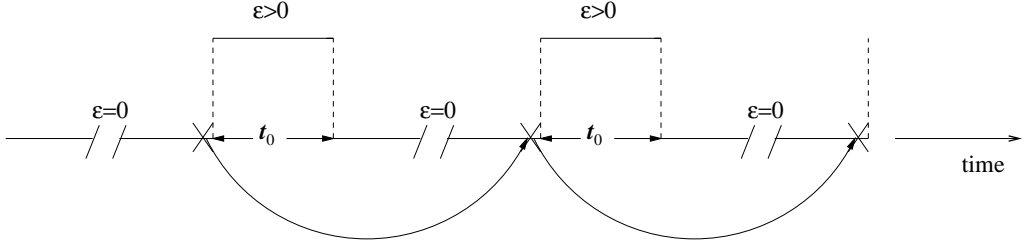}}
\caption{\label{f3}The effective dynamics tries to define the transitions
between the crosses, which correspond to the results of the successive
taps.}
\end{figure}

The stochastic model formulated in the previous section, with the transition
rates given by Eq.\ (\ref{3.1}), can be used to study a variety of compaction
processes by specifying the time dependence of the control parameter
$\epsilon$. In particular, we have already discussed some applications to
tapping processes, based on the particularization of the general equations
following from the master equation to a specific way of vibrating the system.
A different, and more appealing, approach is to look for an effective
master equation following from (\ref{3.1}) and being appropriated for a given
experiment. We have developed such a program for tapping processes
\cite{BPyS00a}. The idea is to look for effective transitions rates,
$W_{ef}({\bf m}|{\bf m}^{\prime})$, connecting the initial and final states
of the system when it is submitted to a
``elementary event'' (see Figure \ref{f3}). The latter is defined as the
combination of a tap and the posterior free relaxation to a metastable
configuration. In the limit of $\epsilon \nu t_{0} \ll 1$, three groups
of possible transitions are identified: \\
a){\em Elementary diffusive events}, conserving the number of particles. They
correspond to the interchange of a hole and a particle,
\begin{equation}
\label{4.1}
\ldots 100 \ldots \rightarrow  \ldots 010 \ldots,   \\
\ldots 001 \ldots \rightarrow  \ldots 010 \ldots.
\end{equation}
The effective transition rate for each of these processes is $\alpha/2$, where
\begin{equation}
\label{4.2}
\alpha= \frac{\nu t_{0} \epsilon}{2}.
\end{equation}
is a positive constant playing a role similar to that played by $\Gamma$ in
the real experiments. \\
b){\em Transitions increasing the number of particles}. There are three of them,
\begin{equation}
\label{4.3}
\ldots 101 \ldots \rightarrow \ldots 010 \ldots ,
\end{equation}
with transition rate $\alpha/2$,
\begin{equation}
\label{4.4}
\ldots 101 \ldots \rightarrow \ldots 001 \ldots,
\end{equation}
with transition rate $\alpha/4$, and
\begin{equation}
\label{4.5}
\ldots 101 \ldots \rightarrow \ldots 100 \ldots,
\end{equation}
also with transition rate $\alpha/4$. \\
c){\em Transitions decreasing the number of particles}. These are
\begin{equation}
\label{4.6}
\ldots 00100 \ldots \rightarrow \ldots 01010 \ldots,
\end{equation}
with transition rate $\alpha^{2}/2$ and
\begin{equation}
\label{4.7}
\ldots 01000 \ldots \rightarrow \ldots 01010 \ldots,
\end{equation}
\begin{equation}
\label{4.8}
\ldots 00010 \ldots \rightarrow \ldots 01010 \ldots,
\end{equation}
both with $W_{ef}=\alpha^{2}/4$. Only those variables corresponding to sites
whose state is changing or conditioning the transition are represented in the
above expressions.  The equivalence of this model to the original one defined by
the transition rates (\ref{3.1}) in the limit of small $\alpha$, has been tested
by comparing the Monte Carlo simulation results obtained with both models. In
fact, the data show that, for the density relaxation, the results from the
effective model and the original one differ by less than $2$ per cent for
$\alpha\leq 0.5$. For constant $\alpha\neq 0$, the system evolves from the
initial low density configuration to a final state characterized by a density
\begin{equation}
\label{4.9}
\rho_{s}(\alpha)=\frac{1}{2} \left[ 1+\left( 1+4\alpha \right)^{-1/2}
\right].
\end{equation}

\begin{figure}
\centerline{\includegraphics[scale=0.5]{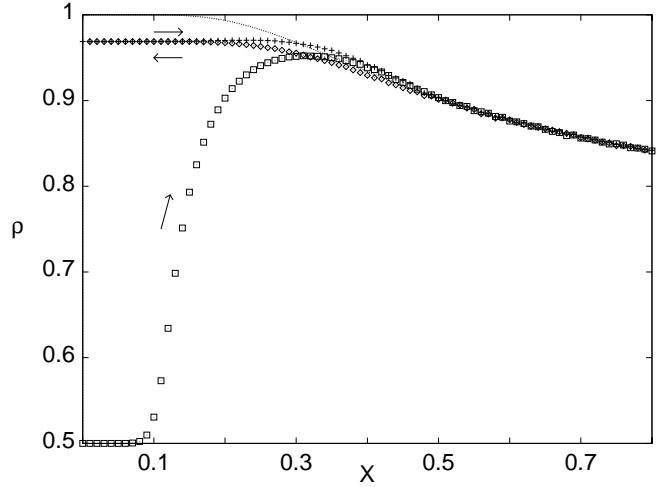}}
\caption{\label{f4}Evolution of the density when the system is submitted to
a tapping cycle as described in the text. The diamonds and the crosses represent
the approximately reversible cooling and heating processes, respectively.}
\end{figure}

A detailed analysis of the properties of the steady state
\cite{BPyS00a} shows that they are consistent with Edwards's theory
\cite{EyO89,MyE89,EyM94}, with the compactivity being identified as $X=
-(\ln \alpha)^{-1}$. Furthermore, when the system described by the effective
transition rates is submitted to processes in which the tapping intensity is
first monotonically increased and afterwards decreased also monotonically, its
time evolution presents the reversible-irreversible branches observed in
experiments \cite{PByS00}. In Figure \ref{f4} an example of the response of the
system to one of these cycles is presented. The compactivity $X$ of the system
is decreased and increased with the same rate, $r=10^{-5}$. Starting from the
loosest configuration, for large enough vibration intensity (or compactivity)
the density of the system approaches the steady curve. Afterwards, when the
compactivity is decreased and increased, always with the same rate, two
(approximately) reversible curves are obtained. These hysteresis effects are
related with the existence of a ``normal evolution curve'', fully determined by
the program of increase of the intensity, and having the strong property of
attracting any other solution of the master equation with the same program
\cite{PByS00}.

\section{Memory effects}
\label{s5}
\begin{figure}
\centerline{\includegraphics[scale=0.5]{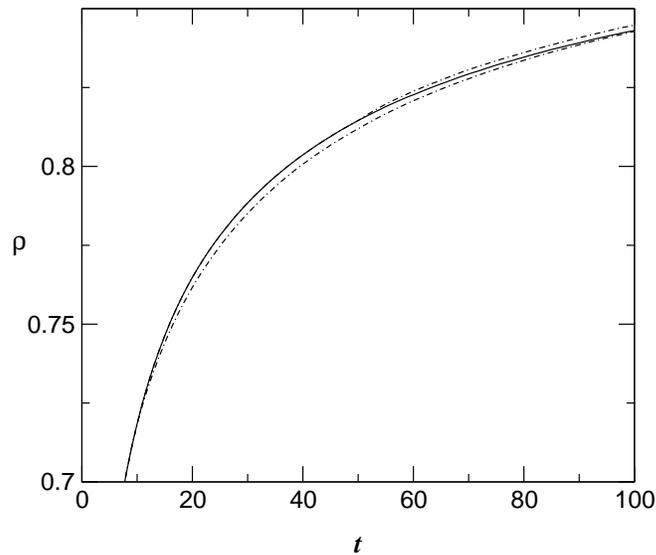}}
\caption{\label{f5}Response of the system to a sudden change in the vibration
intensity from $\alpha=0.15$ to $\alpha=0.125$. When the jump is introduced
at $t_{w}=10<t_{c}$, the response is normal, i.e. the slope of the
relaxation curve decreases. On the other hand, for $t_{w}=50>t_{c}$ an
anomalous response is observed.}
\end{figure}

Now, we will show that the effective dynamics introduced above leads to
a model for tapping processes that belongs to the general class discussed
in Section \ref{s2} and, consequently, also presents the short term memory
effects observed in vibrated granular materials. We start by noting that
the steady density for constant intensity, given by  (\ref{4.9}) is
a monotonic decreasing function of $\alpha$, going from $\rho_{max}=1$
to $\rho_{min}=0.5$. Next, from the master equation with the transition
rates (\ref{4.1})-(\ref{4.8}) it is obtained
\begin{equation}
\label{5.1}
\frac{d\rho}{dt}= \alpha x_{101}(t)- \frac{\alpha^{2}}{2}
\left[ x_{00100}(t)+\frac{1}{2}x_{01000}(t)+\frac{1}{2}x_{00010}(t) \right],
\end{equation}
where $x_{101}$ is the concentration of hole-particle-hole clusters,
$x_{00100}$ is the concentration of two particles-hole-two particles clusters,
and so on. Comparison of the above equation with  (\ref{2.7}) shows that
both equations have the same form, with the choices
\begin{equation}
\label{5.2}
f_{1}(\alpha)=g(\alpha)=\alpha,
\end{equation}
\begin{equation}
\label{5.3}
\mu_{1}(t)=x_{101}(t), \quad
\mu_{2}(t)=\frac{1}{2} x_{00100}(t)+\frac{1}{4} \left[ x_{01000}(t)+
x_{00010}(t) \right].
\end{equation}
Since $f_{1}$ and $g$ are monotonic increasing functions of $\alpha$,
and $\mu_{1}$ and $\mu_{2}$ are defined positive quantities, the model
verifies the conditions required by the validity of the discussion carried
out in Section \ref{s2}. Therefore, we can directly write from Eq.
(\ref{2.10})
\begin{equation}
\label{5.4}
\frac{\Delta r_{w}}{\Delta \alpha}=\lambda(t_{w}),
\end{equation}
where the function $\lambda (t)$ determining the nature of the response of
the system is
\begin{equation}
\label{5.5}
\lambda (t)=\frac{r(t)}{\alpha}-\alpha \mu_{2}(t).
\end{equation}
For instance, for $\alpha=0.15$, Monte Carlo simulations show that $\lambda
(t)>0$ for $t<t_{c}$, while $\lambda (t) <0$ for $t>t_{c}$, with $19 < t_{c}<
20$ \cite{ByP01a}. According to the theory presented here, when the vibration
intensity $\alpha$ is modified at a time $t_{w}<t_{c}$, a normal response in
which the intensity jump and the relaxation rate jump have the same sign is to
be expected. On the other hand, when $t_{w}>t_{c}$, stimulus and response should
have opposite signs, corresponding to an anomalous response. In Figure \ref{f5}
we have plotted the time evolution of the density of a system which is being
vibrated with $\alpha=0.15$, and the intensity is suddenly decreased to
$\alpha=0.125$ at $t=t_{w}$. The only difference between the two curves is in
the value of $t_{w}$; in one case it is $t_{w}=10 <t_{c}$ while in the other
$t_{w}=50>t_{c}$. In agreement with the theoretical analysis, the relaxation
rate at the jump decreases in the first case and increases in the latter.

It is important to stress the generality of the arguments  in
Section \ref{s2}. Although we have restricted ourselves in this work to
a particular simple model of compaction, the theoretical scheme presented
there is rather general. For instance, an equation like (\ref{2.7}) is also found
for the so-called parking model \cite{BKNJyN98,KyB94}. Even more, the
one-dimensional Ising model with Glauber dynamics also belongs to the group
\cite{ByP01b}. In summary, the memory effects discussed in the context
of compaction in granular materials seem to be quite general effects
showing up in many different systems.

\ack
Partial support from the Direcci\'{o}n General de Investigaci\'{o}n
Cient\'{\i}fica y T\'{e}cnica (Spain) through Grant No. PB98-1124 is
acknowledged.


\begin{thebibliography}{99}
\bibitem{note1}
The insert in Figure 5 of Reference \protect{\cite{KFLJyN95}} can be
 misleading, since the parameter $\rho_{\infty}$ shows there an increasing
 behaviour with $\Gamma$. Nevertheless, this is due to the fact that the
 states considered for small values of $\Gamma$ do not correspond to
 the final stable configuration, but to metastable ones
 \protect{\cite{NKBJyN98}}
\end{thebibliography}
\end{document}